\documentclass[reprint,
superscriptaddress,
amsmath,amssymb,
aps,
prl,
twocolumn,
floatfix,]{revtex4-1}
\usepackage{afterpage}
\usepackage{graphicx}
\usepackage{graphics}
\usepackage{wrapfig}
\usepackage{dcolumn}
\usepackage{amssymb}
\usepackage{bm}
\usepackage{epsfig}
\usepackage{float}
\usepackage{psfrag}
\usepackage{color}
\usepackage{ucs}
\usepackage{amsmath}
\usepackage{amsfonts}
\usepackage{amssymb}
\usepackage{graphicx}
\usepackage{dcolumn}
\usepackage{bm}
\usepackage{latexsym}
\usepackage{xr-hyper}
\usepackage[breaklinks=true]{hyperref}
\usepackage{breakcites}
\usepackage{xcolor}
\newcommand{\HL}[1]{\textcolor{black}{#1}}

\hypersetup{
    colorlinks,
    linkcolor={purple!50!black},
    citecolor={purple!50!black},
    urlcolor={purple!80!black}
}

\begin{document}
\title{Bulk condensation 
by an 
active 
interface
} 
\author{Raushan Kant}
\email{raushankant@iisc.ac.in}
\affiliation{
Department of Physics, Indian Institute of Science, Bangalore 560 012, India}
\author{Rahul Kumar Gupta}
\email{rahul.gupta09041@gmail.com}
\affiliation{Sankhyasutra Labs Ltd., Manyata Embassy Business Park, Bengaluru, Karnataka 560 045, India}
\author{Harsh Soni}
\email{harsh@iitmandi.ac.in}
\affiliation{School of Physical Sciences, Indian Institute of Technology Mandi, Kamand 175 005, India}

\author{A K Sood}
\email{asood@iisc.ac.in}
\affiliation{Department of Physics, Indian Institute of Science, Bangalore 560 012, India}
\author{Sriram Ramaswamy}
\email{sriram@iisc.ac.in}
\affiliation{
Department of Physics, Indian Institute of Science, Bangalore 560 012, India}

\date{\today}

\begin{abstract} We present experiments, supported by mechanically detailed simulations, establishing bulk 
condensation of a hard-bead fluid by a tiny population of orientable motile grains that self-assembles into a moving polarized monolayer. In a quasi-1D geometry two such layers, oppositely aligned, immobilize the condensed non-motile component. We account for our observations through a continuum theory with a naturally non-reciprocal Cahn-Hilliard structure, whose predicted trends as a function of packing fraction are consistent with our observations.

\begin{description}

\item[Subject areas]
Physical Systems: Living matter \(\&\) active matter 
\end{description}
\end{abstract}

\maketitle

The persistent motility of active matter \cite{ReviewSriram2010,ReviewMarchettiSriram2013} leads to a panoply of striking behaviours \cite{VICSEK201271,PhysRevX.12.010501,ReviewBechinger2016,chate2020dry,gompper2022active}, of which condensation with neither attraction or depletion \cite{tailleur2008statistical,Fily2012, McCandlish2012,Redner2013,Cates2015,geyer2019freezing} and motile-nonmotile demixing \cite{Stenhammar2015,wysocki2016propagating,Dolai2018,agrawal2021alignment} are the focus of the present work. 
In this article we report unexpected states of organisation in experiments, supported by mechanically detailed simulations, on mixtures of motile polar aligning rods and non-motile spherical beads. In dimensions $d=1$ and $2$, we find spontaneous segregation of polar rods and beads with increasing area fraction of either or both species. 

\begin{figure}
    \centering
    \includegraphics[width=0.48\textwidth]{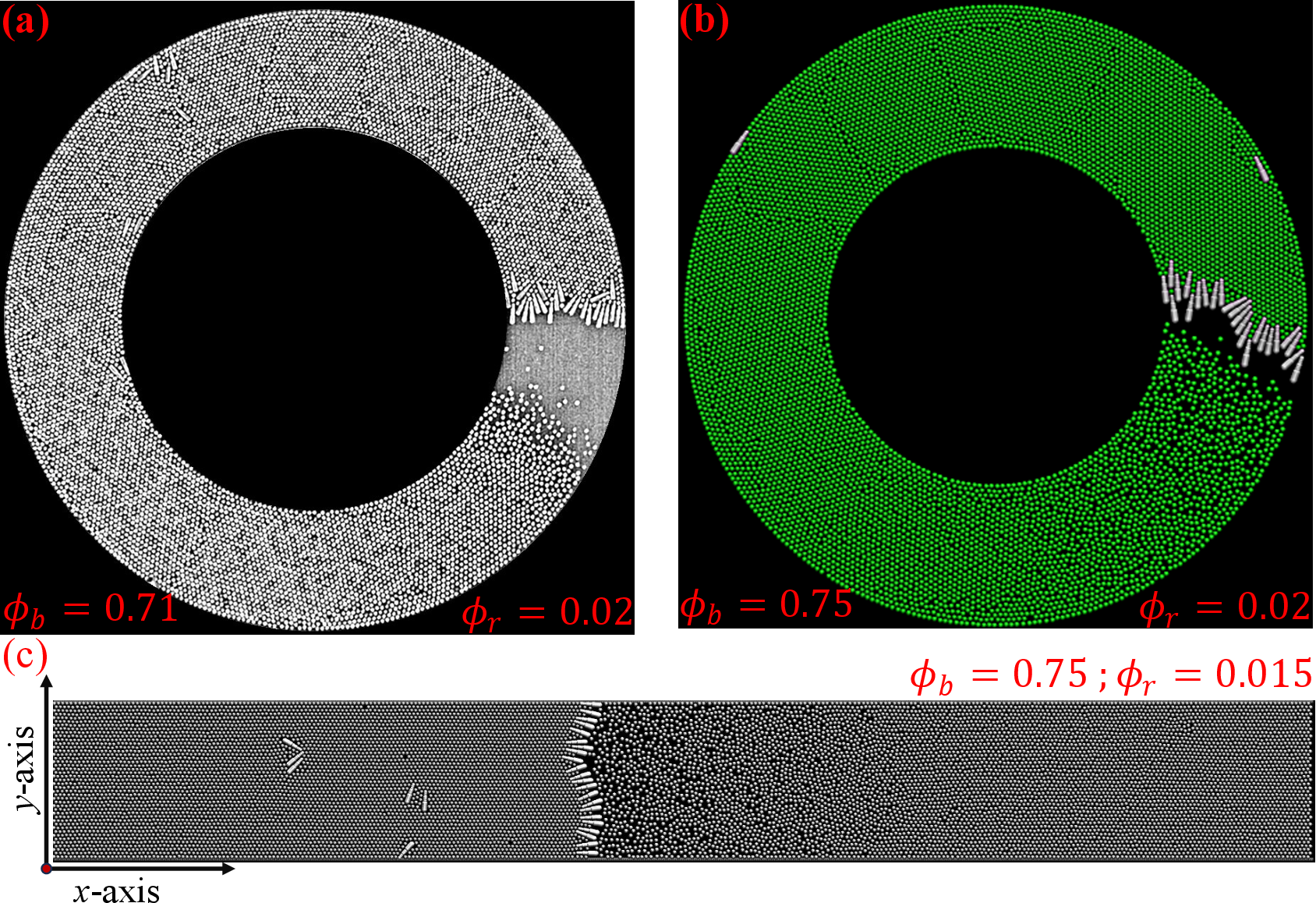}
    \caption{A self-assembled monolayer of polar active particles condenses a bulk quantity of passive beads in (a) experiments and (b) mechanically detailed simulations on vibration-activated granular matter in an annular channel. The rod area fraction \(\phi_{r} = 0.02\), bead area fraction \(\phi_{b} = 0.71\) in experiment and \(\phi_{b} = 0.75\) in simulation, if averaged over the entire domain. (c) The same phenomenon realized in simulations in a straight channel with dimensions $249 \times 31$ in units of the bead diameter, with periodic boundary conditions in the long dimension, at bead and rod area fractions \(\phi_{b} = 0.75\) 
    \(\&\) \(\phi_{r} = 0.015\).}
    \hypertarget{self-assembled-fig}{}
    \label{self-assembled}
    
\end{figure}

\begin{figure*}
    \centering
    \includegraphics[width=1\linewidth]{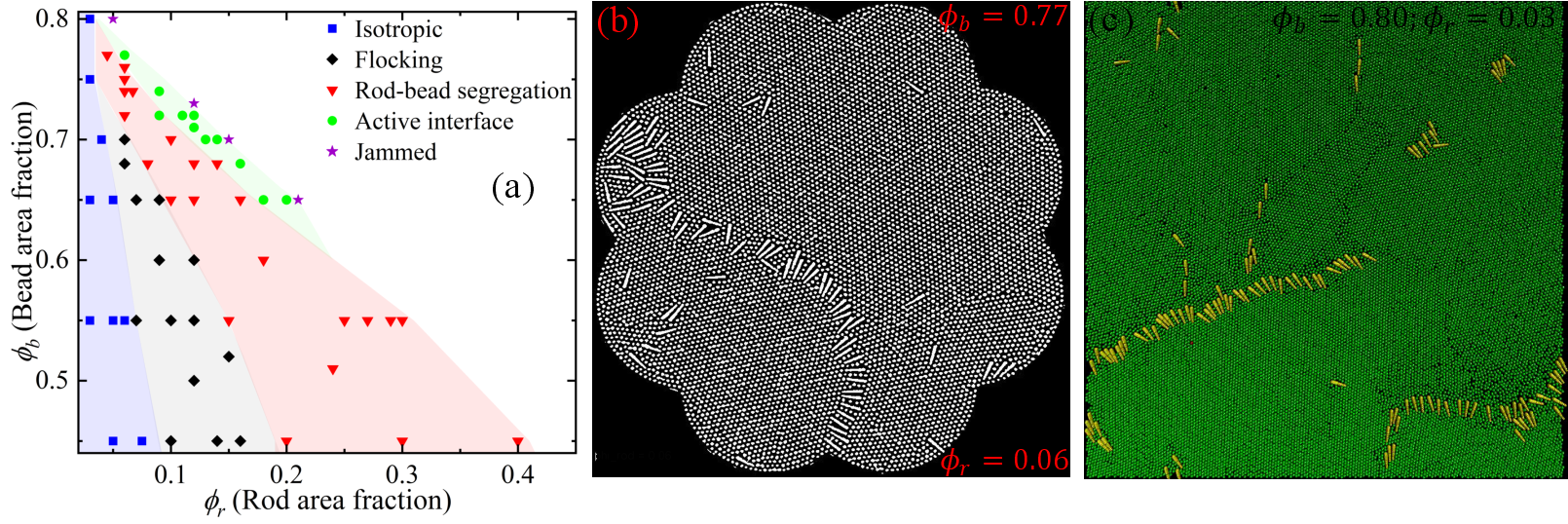}
    \caption{(a) Experimental phase diagram in 2D in \(\phi_{b}\)-\(\phi_{r}\) plane, showing isotropic phase, flock, rod-bead segregation, and condensation by active interface,
    also see movie SM6 \cite{MOV}. ``Jammed'' in the figure refers to a state in which extreme packing (total area fraction \(\simeq 0.85\)) leads to nearly arrested dynamics. (b) Bulk condensation of beads by an active interface of aligned rods in the experiment; \(\phi_b = 0.77\) and \(\phi_r = 0.06\). (c) Travelling interface of aligned rods in a simulation with periodic boundary conditions in both directions of a square box of length 249 bead diameters; \(\phi_b = 0.80\) and \(\phi_r = 0.03\).}
    \hypertarget{ds}{}
    \label{ds}
\end{figure*}

The phenomenon persists down to exceptionally low rod fraction: a single row of motile polar rods is able to condense a bulk domain of beads, see Fig. \hyperlink{self-assembled-fig}{\ref{self-assembled}}. Thus a finite number in $d=1$, and a subextensive fraction in $d=2$, of motile rods apparently affects an extensive population of beads. In $d=2$, the rods self-assemble into a single row moving transverse to its length, forming a one-dimensional polarized active membrane, condensing the beads ahead of it (Fig. \hyperlink{ds}{\ref{ds}}, movies SM1 \(\&\) SM2 \cite{MOV}). In an annular channel with width much smaller than circumference, effectively $d=1$ (Fig. \hyperlink{self-assembled-fig}{\ref{self-assembled}}, movies SM3 \(\&\) SM4 \cite{MOV}), the active membrane is especially robust and width-spanning. Upon increasing the rod fraction, two membranes with opposing polarity self-assemble to immobilize a macroscopic bead-dense domain, Fig. \hyperlink{self_trapping}{\ref{self_trapping}}(a), movie SM5 \cite{MOV}. 
The segregation can be understood physically through a positive-feedback argument: the moving rods force a bead-density gradient in the direction of their motion; their orientation is sterically adapted to point up this gradient.
This mechanism operates through a coarse-grained theory 
for the dynamics of rod and bead densities, which naturally takes a non-reciprocal Cahn-Hilliard form \cite{saha2020,you2020,saha2208effervescent,saha2024phase,tucci2024nonreciprocal,greve2024maxwell}. Finally, we show theoretically that a localized motile rod density can transport a bead domain whose extent grows with its compression modulus, rationalizing the macroscopic condensation of beads. \HL{In an interesting contrast with \cite{Wang2023, Derivaux2023, Stengele2023capture}, the enhanced transport in our case is thanks to organizing the \textit{cargo}.} The mechanisms we describe are generic, relying only on motility, shape, and excluded volume, and thus should arise in living matter.  

We now show how we obtained these results.  
Our experimental system consists of brass rods of length $4.5$ mm, with thickness tapered from $1.1$ mm to $0.7$ mm from tail to nose, and spherical aluminium beads~\cite{narayan2010phase, kumar2014flocking, Kumar2019, Bera2020, Soni2020, gupta2022active} of diameter \(1\) mm. The rod-bead mixture is confined between an anodised aluminium plate of diameter \(12\) cm and a glass lid, whose surfaces are separated by \(1.12\) mm. The plate is attached to a permanent magnetic shaker (LDS V406-PA100E). All experiments are done at a shaker frequency \(f=200 \) Hz and amplitude \(\mathcal{A}=0.04\) mm, corresponding to a shaking strength \(\left(2\pi f\right)^2\mathcal{A}=\) $7$ times the acceleration due to gravity.

We create an annular geometry by glueing a circular block at the centre of the sample cell; see Fig. \hyperlink{self-assembled-fig}{\ref{self-assembled}}(a). We capture images at six frames per second on a Redlake MotionPro X3 camera and post-process them on Fiji (ImageJ) for analysis in MATLAB and Python.
We present complementary results from granular dynamics simulations as in \cite{kumar2014flocking, Soni2020,Sharma2024Phases}, with inter-particle and particle-wall interactions captured by an impulse-based collision model~\cite{stronge1994}, including static friction and finite restitution \HL{which we can tune to change the motility of the polar rods}.
We use periodic boundary conditions (PBC) in the $xy$-plane to eliminate the complicating effect of lateral walls. In some cases we also re-create the experimental geometry~\cite{Soni2020} by way of confirmation. We normally set the static friction and restitution coefficients to 0.05 and 0.3 for particle-particle collisions, 0.03 and 0.1 for rod-base and rod-lid collisions, and 0.01 and 0.3 for bead-base and bead-lid collisions, respectively, in reasonable correspondence with the experimental system~\cite{kumar2014flocking, soni2019flocks, Soni2020}. We have established the stationarity of our experimental nonequilibrium phases for at least 300 and, in several cases, 750 seconds. We use VMD software \cite{VMD} to make simulation movies and snapshots. 

For rod area fraction $\phi_r=0$ the system displays fluid, hexatic and crystalline phases as a function of bead area fraction $\phi_b$ \cite{gupta2022active}. Prior studies of motile polar rods introduced into this system found flocking through a flow-induced interaction \cite{kumar2014flocking}, and non-reciprocal pair interactions in crystalline \cite{gupta2022active} and dense fluid backgrounds \cite{banerjee2022unjamming}. Fig. \hyperlink{ds}{\ref{ds}}(a) shows our experimental phase diagrams in the \(\phi_b-\phi_r\) plane. For $d=2$, at \(\phi_b = 0.55\) and \(\phi_r = 0.06\), the system is in the isotropic state, with particles distributed homogeneously and without organised motion or alignment. Across a first threshold (\(\phi_b =0.70\) and \(\phi_r =0.06\)), we see a transition to a homogeneous flock as in \cite{kumar2014flocking} with polar-rod orientations and all particle velocities aligned along the azimuthal direction. Past a second threshold (\(\phi_{b} = 0.72\) and \(\phi_r =0.06\)), we observe rod-bead segregation. The third and most striking threshold ($\phi_b = 0.77$ and \(\phi_r =0.06\)), marks the formation of a self-assembled row -- a polar monolayer -- of active rods, which condenses a bulk domain of beads [see Fig. \hyperlink{ds}{\ref{ds}}(b) and movie SM6 \cite{MOV}]. The rods align 
side by side, with their tapered tips pointing towards the dense, immobile region. The locus of the onset of this active self-assembly is $\phi_r+\phi_b \approx 0.83$. The bead domains on 
the two sides of the monolayer differ distinctly, though slightly, in packing fraction and considerably in mobility; see Fig. S1 \cite{SUPP} \nocite{ramaswamy2000nonequilibrium,turlier2019unveiling,del2019interface,patch2018curvature,gokhale2013grain,brotto2013hydrodynamics,olafsen2005two} and movie SM1 \cite{MOV}. 
Height fluctuations of the monolayer consistent with the existence of an interfacial tension are seen over a limited dynamic range, see Fig. S4 \cite{SUPP}. \HL{The ratio of energy scale to tension implied by our measurements yields \cite{SUPP} a length $1.90 \pm 0.05$ mm, comparable to our particle size, and consistent with such a correspondence in a quite different granular system \cite{clewett2012emergent}.} 
In simulations in a square domain with PBC, we find coherently moving self-assembled 1D membranes made of polar rods pointing from low to high density at \(\phi_b = 0.80\) \(\&\) \(\phi_r = 0.03\) [see Fig. \hyperlink{ds}{\ref{ds}}(c) and movie SM2 \cite{MOV}]. \HL{We have carried out simulations over a range of $\phi_b$ in rectangular domains \textit{without} PBC as well, and again find coexistence between regions of reproducible high and low bead density (see movie SM13 \cite{MOV}). 
Varying the single-rod motility $u_0$, we establish a threshold below which phase separation is absent, see SM12 \cite{MOV}; for intermediate $u_0$, phase separation with well-defined nonzero coexistent densities on both sides of the motile-rod layer is seen, see Fig. S4. The shock-like feature in the density is observed only at large enough $u_0$, see Fig. S4 \(\&\) movie SM12 \cite{MOV}}. Thus, our experiments and simulations allow an exceptionally small (and possibly sub-extensive) fraction of active rods to condense a bulk domain of passive beads. 

Figs.\hyperlink{self-assembled-fig}{\ref{self-assembled}}(a) and (b) show that the phenomenon persists in experiments and simulations in the quasi-1D setting of an annular channel. For \(\phi_{b} = 0.71\) and \(\phi_{r} \simeq 0.02\), the system transitions from the isotropic phase to a state in which a self-assembled membrane composed of just enough rods to fill a channel width separates bead-rich and bead-poor domains, see SM7 \cite{MOV}.
Rods initially dispersed throughout the medium in an isotropic state spontaneously undergo a high degree of spatial localization and azimuthal alignment. Fig. S2(b) shows that the bead medium is close-packed and nearly incompressible in front of the self-assembled active layer and exceedingly small behind it \cite{SUPP}. Fig. \hyperlink{self-assembled-fig}{\ref{self-assembled}}(c) from our simulations in a quasi-1D geometry with PBC along the \(x\) axis shows the formation of a self-assembled active membrane at large enough $\phi_b$, reinforcing our experimental observation without the complication of a curved boundary. The membrane persists in simulation upon doubling the long dimension of the box with all other parameters unchanged; see SM8 \cite{MOV}. We confirm the stationarity of the self-assembled state for over 2000 seconds; see SM3 \cite{MOV}. In fact the sequence of phases -- disordered to flock to condensation by active interface -- with increasing \(\phi_{b}\) at a given \(\phi_{r}\) is seen in the quasi-1D system as well, [see Fig. \hyperlink{self_trapping}{\ref{self_trapping}}(a) and movie SM9 \cite{MOV}].  

The most dramatic experimental demonstration of bulk condensation by an active boundary is seen in the quasi-1D geometry when the number of rods is sufficient to form two spanwise rows. In an ordered and segregated configuration, a subset of motile rods defects from the trailing to the leading edge of the moving bead domain, capturing and immobilizing it permanently on the time scale of our experiment, [see Figs. \hyperlink{self_trapping}{\ref{self_trapping}})(b), (c) and movie SM5 \cite{MOV}]. 
Imposing a scalloped periphery \cite{deseigne2010collective} on the annular geometry facilitates the turning of the rods, augmenting this active capture, see Fig. \hyperlink{self_trapping}{\ref{self_trapping}}, and leads to a stronger density enhancement in the condensed region, see Fig. S3 for characterisation of bead medium \cite{SUPP}. 
A self-sorting mechanism accounts for the stability of the balanced, immobile state: unequal rod populations at the two ends of a bead condensate will lead to net unidirectional motion, liberating rods at the trailing edge, which can turn and eventually be captured by the leading edge. 

We note that active capture can be viewed as a 1D version of the core-halo state observed in motile/non-motile mixtures in \cite{Stenhammar2015}. In two dimensions, the steric interaction between our elongated motile particles promotes alignment, leading to extended domains with nearly straight boundaries rather than the irregular rafts of \cite{Stenhammar2015}.
\afterpage{
  \begin{figure*}
    \centering
    \includegraphics[width=1\textwidth]{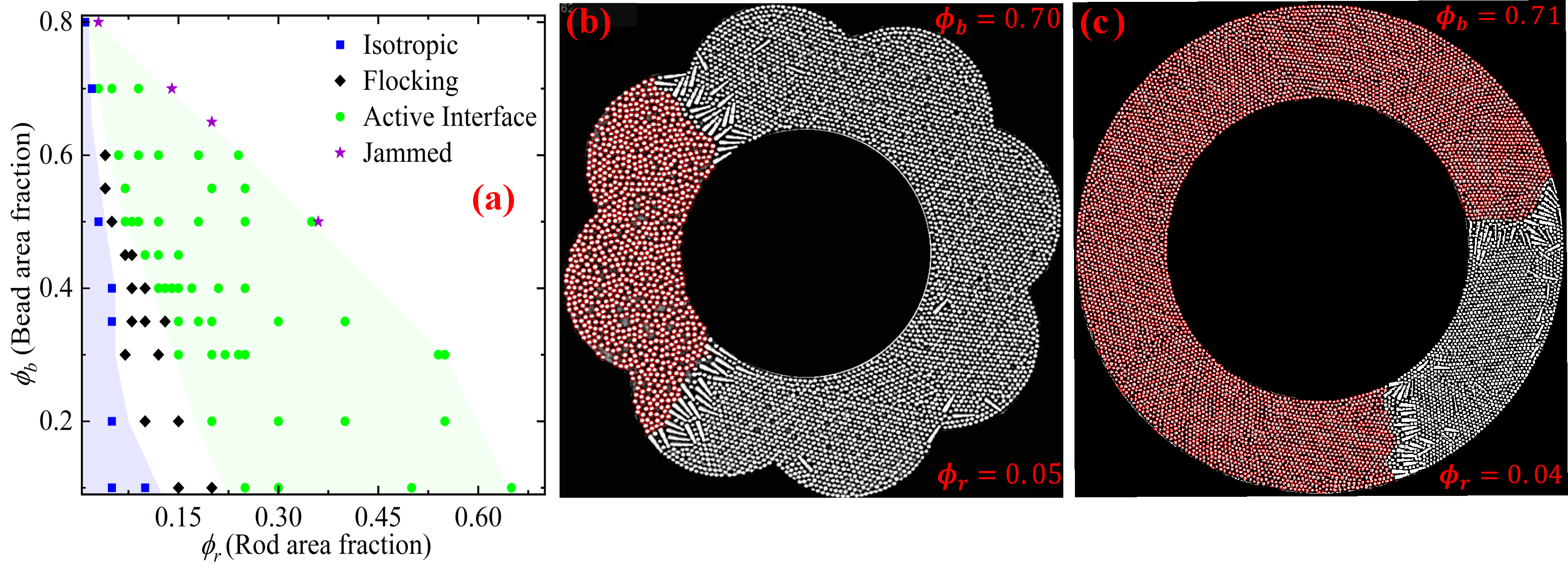}
    \caption{ (a) Phase-diagram in quasi-1D in \(\phi_r\) - \(\phi_b\) plane, which shows series of phases: isotropic, flocking to condensation by active interface,
    also see movie SM9 \cite{MOV}. (b) and (c) shows a snapshot of condensation by two monolayers of rods in quasi-1D geometry with and without the floral outer boundary, with \(\phi_b = 0.70\), \(\phi_r = 0.05\) and \(\phi_b = 0.71\), \(\phi_r = 0.04\) respectively. The radius of the inner block in (b) and (c) is 2.5 and 4 cm respectively.} 
    \hypertarget{self_trapping}{}
    \label{self_trapping}
\end{figure*}
}
We now show that the mechanism and broad features of the demixing of active rods and passive beads can be understood within a coarse-grained theory. The hydrodynamic fields in our description are the number densities $\rho$ and $\sigma$ of beads and rods, obeying continuity equations $\partial_t \rho + \mathbf{\nabla}\cdot \mathbf{J}_{\rho} =  0, \, \partial_t \sigma + \mathbf{\nabla} \cdot \mathbf{J}_\sigma  =  0$, with currents $\mathbf{J}_{\rho} =  - D_{\rho} \nabla \rho -  D_{\rho \sigma}\nabla \sigma + \alpha \rho \mathbf{P}, \, \mathbf{J}_{\sigma}  = - D_{\sigma} \nabla \sigma - D_{\sigma \rho}\nabla \rho + u_0 \sigma \mathbf{P}$, where the bare diagonal diffusivities $D_{\rho}, D_{\sigma}$ and cross-diffusivities $D_{\rho \sigma}, D_{\sigma \rho}$ govern the \textit{passive} dynamics (and hence $D_{\rho \sigma} D_{\sigma \rho} \ge 0$). Self-propulsion of rods with speed scale $u_0>0$, and active forcing of beads by rods with strength $\alpha>0$ \cite{kumar2014flocking}, are guided by the vector orientational order parameter field $\mathbf{P}$ -- hereafter the polarization -- which microscopically is the local average of unit vectors from the broad to the narrow end of the polar rods. We take $D_{\rho},\, D_{\sigma}>0$, as we are not in a regime where we expect either component to condense in the absence of interactions. \HL{In the isotropic phase, it suffices to assume $\mathbf{P}$ relaxes rapidly to a profile determined by anchoring to density gradients: $ (a - K \nabla^2) \mathbf{P} = A \mathbf{\nabla} \rho + B \mathbf{\nabla} \sigma$, with phenomenological couplings $a (>0\, \mbox{in the isotropic phase}), \, A,B$ and an elastic stiffness $K$. See \cite{SUPP} for a more detailed analysis, including the effect of self-advection of $\mathbf{P}$ \cite{toner1998,TONER1995} and other minor modifications to this picture for dynamics in the ordered phase. 
Perturbing $\rho = \rho_0 + \delta \rho, \, \sigma = \sigma_0 + \delta \sigma$ deep in the isotropic phase ($a$ large and positive) yields, to $4$th order in gradients, the linear stability equations}
\begin{widetext}
\begin{equation} 
\label{rhoeqiso}
\partial_t \delta \rho = 
\left[\left(D_{\rho} - {\rho_0 \alpha A \over a}\right) 
\nabla^2
-  {\rho_0 \alpha A K \over a^2} 
\nabla^4 \right]\delta \rho
+ \left[\left(D_{\rho \sigma}
-{\rho_0\alpha B \over a}\right) 
\nabla^2
-{\rho_0\alpha B K\over a^2} 
\nabla^4 \right] \delta \sigma, 
\end{equation}

\begin{equation} 
\label{sigmaeqiso}
\partial_t \delta \sigma = \left[\left(D_{\sigma \rho}  - {\sigma_0 u_0 A \over a}\right) 
\nabla^2
- {\sigma_0 u_0 A K  \over a^2} 
\nabla^4\right]
 \delta \rho + \left[\left( D_{\sigma} - {\sigma_0 u_0 B \over a}\right)  
 \nabla^2 - {\sigma_0 u_0 B K \over a^2} 
 \nabla^4\right]\delta \sigma.  
\end{equation}
   \end{widetext}
\HL{In \eqref{rhoeqiso} and \eqref{sigmaeqiso}, a large positive $\rho_0 \alpha A / a$ would lead to both bead condensation, in the form of a diffusively growing mode, and  
stabilizing ``surface tension'' terms at order $\nabla^4$. 
There is a nice parallel here with phase separation in equilibrium \cite{cahn1958free,bray1994theory,chaikin1995principles}, where both these classes of terms would have a single microscopic origin, namely, an attractive interaction. We now offer a physical argument for the positivity of $\alpha A>0$, elaborating on the feedback argument presented at the start of this article.}
Sterically, a region of high bead density will accommodate a polar rod's narrow nose more readily than it will its broad rear. Statistically, therefore, polar rods are likely to be found pointing from low to high bead density, so $A>0$. Rods move with their narrow end forward, and propel the beads by contact. This creates $\rho$ gradients in the direction of $\mathbf{P}$, hence $\alpha > 0$. Moreover, the independent activity parameters $\alpha$ and $u_0$ can cause beads to flee high rod density and rods to pursue high bead density, leading to opposite signs for the off-diagonal couplings in \eqref{rhoeqiso} in \eqref{sigmaeqiso}. This realizes the non-reciprocal Cahn-Hilliard system of \cite{saha2020,you2020}, with the added feature that even the phase separation is a consequence of activity. In the simplifying limit \(D_\rho = 0\), \(D_\sigma = 0\) \(\&\) \(D_{\sigma\rho} = 0\), the diffusive instability, that is bead condensation, arises for \(A \alpha \rho_0+B u_0 \sigma_0 > 0\), and non-reciprocity in the sense of off-diagonal couplings with opposite signs for \((A \alpha \rho_0 - B u_0 \sigma_0)^2 + 4AB\alpha\rho_0u_0\sigma_0 < 4aA D_{\rho\sigma} u_0 \sigma_0\). These two conditions can coexist, e.g., for large enough $D_{\rho\sigma}$. More generally, decreasing $D_{\sigma\rho}$, say by increasing rod length, keeping other parameters fixed favors the emergence of non-reciprocity. When both instability and non-reciprocity are present, traveling spinodal patterns should arise as in \cite{saha2020,you2020}, a direction that remains to be explored in our experiments. Calculational details and the equations for perturbations deep in the ordered phase ($-a = |a|$ large), where non-reciprocal effects enter at sub-leading order in wavenumber, are in the SI \cite{SUPP}, with  complete expressions for mode frequencies in Eq. (S10) and (S11).

Having established the mechanism for bulk rod-bead segregation, we finally return to the 
condensation of a \textit{bulk} domain of beads by an \textit{interfacial} population of motile rods. It suffices to examine the problem in one dimension. Details are in the Supplement \cite{SUPP}. For a closely related calculation see \cite{banerjee2022unjamming}. We replace the rods by a point force density $f \delta(x-v_0 t)$ moving with speed $v_0$ \HL{(which we expect to scale with $u_0$ in \eqref{sigmaeqiso})} in the $x$ direction, in a bead medium with density $\rho$ and current $J_{\rho}$, obeying force balance 
\begin{equation} 
\label{eq:forcebal} 
\gamma J_{\rho} = -\partial_x \Pi(\rho(x)) + f \delta(x-v_0 t)
\end{equation} 
with a drag coefficient $\gamma$ per bead and a general bead-pressure equation of state $\Pi = \Pi(\rho)$, and boundary conditions $\rho(x = - \infty) = \rho_0, \rho'(-\infty) = 0$. By suitable integration across the Dirac delta we show that 
\begin{equation}
\label{eq:Pijump}
\Pi(\rho_+) - \Pi(\rho_-) = f
\end{equation} 
where $\rho_+, \, \rho_-$ are the densities on either side of the point force. 
If we approximate $\Pi'(\rho) \simeq \Pi'(\rho_0)$ for $x<0$ and $\Pi'(\rho) \simeq \Pi'(\rho_+)$ for $x>0$, we find 
$\rho(x<0) \equiv \rho_0$ and 
\begin{equation}
\label{eq:rhoright}
\rho(x>0) = \rho_0 + (\rho_+ - \rho_0) e^{-x/\ell_+}, \, \ell_+ \equiv {\Pi'(\rho_+) / \gamma v_0},  
\end{equation}
with $\rho_+$ determined by $\Pi(\rho_+) - \Pi(\rho_0) = f$. A detailed solution must go beyond the approximation of a constant $\Pi'(\rho)$ and thus requires knowledge of the bead equation of state. For our purposes it suffices to know, from \eqref{eq:rhoright}, that the extent of the bead domain, that is, the length scale $\ell_+$ of the decay of the density with distance from the rod layer, is set by the stiffness $\Pi'(\rho_+)$. With increasing $f$, $\rho_+$ approaches maximum packing, so it becomes possible to push a bead domain of unlimited extent. Presumably, this is the mechanism at work in Fig. \hyperlink{self-assembled-fig}{\ref{self-assembled}}.

Before closing, we comment on other studies of active-passive segregation. In \cite{Dolai2018} small active particles condense domains of large passive particles through a depletion-like \cite{asakura1954interaction} mechanism. \cite{McCandlish2012} examines the role of alignment of velocities and orientations, and \cite{agrawal2021alignment} considers the effect of particle inertia. Our observations are related to the passive-core/active-halo segregation of \cite{Stenhammar2015,wysocki2016propagating}, but the domains in \cite{Stenhammar2015} are short-lived and limited in size, and the passive bands in \cite{wysocki2016propagating} are of finite width and are traversed by an active-particle flux. The absence of aligning interactions of the motile particles with each other or with density gradients is a key difference with respect to our work. An exception that just came to our notice is \cite{kreienkamp2024nonreciprocal}, in which the non-reciprocal interplay of alignment and density fields plays a central role.

We conclude with a summary of our results and a discussion of implications and open issues. Our experimental study of the nonequilibrium phase diagram of mixtures of shape-polar, motile or active rods with a steric tendency towards alignment and non-motile or passive spherical beads uncovers both bulk active-passive segregation and, more remarkably, bulk 
condensation of the beads effected by the activity of an active interfacial layer or membrane of rods. A particularly striking effect in quasi-one-dimensional annular geometries is the formation of an immobile bulk domain of beads sequestered by two opposing active layers. We present a coarse-grained theory of this active condensation, based on the forcing of beads by motile rods and the steric tendency of rods to point in the direction of increasing bead density. The resulting diffusively growing mode presumably marks the limit of stability of the state of homogeneous bead density. 
The theory also predicts conditions under which the condensation should fail. Although the observed joint arrangement of rods and beads is precisely as implied by the above mechanism, our theory describes only the limit of linear stability, i.e., spinodal decomposition. Our observations, however, appear to be in a regime in which a self-assembled active layer of motile rods arises through nucleation. Further, we show analytically that the forcing generated by a localized active polar region can condense a macroscopic bead-rich domain ahead of it, whose density grows with the strength of the forcing and whose extent is limited only by its compressional stiffness, which in turn grows with the density. A theory taking into account the interplay with flocking and band formation as outlined in \cite{gupta2021simulations}, and bridging the gap between monolayer and bulk concentrations of motile polar rods, is in progress. \HL{The self-assembled polar monolayer, as agency for condensation \textit{and} a source of active pressure at a flat interface, should govern the locus of coexistent densities (Fig. S4) and give rise to an effect analogous to ``uncommon tangents'' \cite{Cates2015}. Our preliminary numerical observations (SM13) also suggest also a dependence of the coexisting densities on the overall mean density, a possibility alluded to in \cite{Cates2015}.} 
\HL{Finally, our discovery that a motile periphery can sequester a non-motile bulk offers an original mechanism for condensation in living matter \cite{julicher2024droplet}; a related effect has recently been reported \cite{Utz_preprint}}.
\begin{acknowledgments} We have benefited from valuable discussions with M Barma, M E Cates, A Maitra, R Mondal, C Nardini, and M Rao. SR thanks the SERB, India for a J C Bose Fellowship, ICTS-TIFR for a Simons Visiting Professorship, the Isaac Newton Institute for Mathematical Sciences, Cambridge, for a Rothschild Distinguished Visiting Fellowship and support and hospitality during the programme \href{https://www.newton.ac.uk/event/adi/}{Anti-diffusive dynamics: from sub-cellular to astrophysical scales} where work on this paper was undertaken, supported by EPSRC grant no EP/R014604/1, and ICTS-TIFR for support and hospitality during \href{https://www.icts.res.in/discussion-meeting/amab2023}{Active Matter and Beyond}.
RK was supported by the UGC, India, AKS acknowledges a National Science Chair Professorship of the DST, Government of India, and HS thanks the SERB, India, for support under grant no. SRG/2022/000061-G.

\end{acknowledgments}
\bibliography{library}
\end{document}